\documentclass[twocolumn,preprintnumbers,amsmath,amssymb,showpacs]{revtex4}
\usepackage{epsfig}
\begin{document}
\setlength{\voffset}{1.0cm}
\title{Divergence of the axial current and fermion density in Gross-Neveu models}
\preprint{FAU-TP3-07/03}
\author{Felix Karbstein\footnote{Electronic address:  felix@theorie3.physik.uni-erlangen.de}}
\author{Michael Thies\footnote{Electronic address: thies@theorie3.physik.uni-erlangen.de}}
\affiliation{Institut f\"ur Theoretische Physik III,
Universit\"at Erlangen-N\"urnberg, D-91058 Erlangen, Germany}
\date{\today}
\begin{abstract}
The divergence of the axial current is used to relate the spatial derivative of the fermion density
to the bare fermion mass and scalar/pseudoscalar condensates in 1+1 dimensional
Gross-Neveu models. This serves as a useful test of known results, to explain simple
features of the continuous chiral model and to resolve a conflict concerning the
assignment of baryon number to certain multi-fermion bound states.
\end{abstract}
\maketitle
In this work we consider the family of Gross-Neveu models \cite{1} with Lagrangians
\begin{equation}
{\cal L}  = {\rm i} \bar{\psi}\partial \!\!\!/ \psi -m_0 \bar{\psi} \psi
+ \frac{g^2}{2}  \left[ (\bar{\psi}\psi)^2 + \lambda (\bar{\psi}{\rm i}\gamma_5 \psi)^2\right] 
\label{D1}
\end{equation}
($\lambda=0,1$) in the large $N$ limit. Flavour indices are suppressed, i.e., $\bar{\psi}\psi=\sum_{k=1}^N \bar{\psi}_k\psi_k$ etc. These
are well-studied self-interacting fermion models in 1+1 dimensions, see \cite{2,3,4} and references therein. For
vanishing bare mass $m_0$ they exhibit discrete ($\lambda=0$) or continuous ($\lambda=1$) chiral  symmetry.
For simplicity we shall refer to the first model as Gross-Neveu (GN) model, the second one as
Nambu--Jona-Lasinio (NJL$_2$) model in two dimensions. 
From the Euler-Lagrange equations we derive the divergence of the (flavor neutral) vector
current $j^{\mu}=\bar{\psi}\gamma^{\mu}\psi$ and axial vector current
$j_5^{\mu}=\bar{\psi} \gamma^{\mu}\gamma_5 \psi$,
\begin{eqnarray}
\partial_{\mu} j^{\mu} & = & 0,
\nonumber \\
\partial_{\mu}j_5^{\mu} & = &   2 \left(m_0- (1-\lambda) g^2\bar{\psi}\psi \right) \bar{\psi} {\rm i}\gamma_5\psi .
\label{D2}
\end{eqnarray} 
The first equation expresses conservation of fermion number in all models. The 2nd equation
reflects the continuous chiral symmetry of the massless NJL$_2$ model (for $m_0=0,\lambda=1$) and
exhibits the source of chiral symmetry violation in those cases where the right hand side is non-zero.
At $m_0 \neq 0, \lambda=1,$ it reduces to the standard PCAC relation. In 1+1 dimensions,
vector and axial vector current are trivially related ($\gamma_5=\gamma^0 \gamma^1$),
\begin{equation}
j_5^0 = j^1, \quad j_5^1 = j^0.
\label{D3}
\end{equation}
Taking an expectation value of Eqs.~(\ref{D2}) in a static configuration (vacuum, multi-fermion bound state, 
dense matter) and using the factorization characteristic for the large $N$ limit together with Eqs.~(\ref{D3}), we find
\begin{eqnarray}
\partial_1 \langle j^1 \rangle & = & 0,
\label{D4} \\
\partial_1 \langle j^0 \rangle &=& 2\left(m_0-(1-\lambda) g^2 \langle \bar{\psi}\psi \rangle \right) \langle \bar{\psi}
{\rm i} \gamma_5 \psi \rangle .
\label{D5}
\end{eqnarray}
Likewise we can take the thermal average of Eqs.~(\ref{D2}) in the grand canonical
ensemble at finite temperature and chemical potential. Relations (\ref{D4},\ref{D5}) then remain valid
provided that the expectation values are interpreted as thermal ones.

Whereas Eq.~(\ref{D4}) seems to be trivially fulfilled in all cases considered below and apparently 
contains little dynamics, Eq.~(\ref{D5}) turns out to be quite powerful. It relates the spatial derivative of
the fermion density
to the scalar and pseudoscalar condensates and the bare mass. In the remainder of this paper we will confront this
identity to what is known about GN models to illustrate its potential use. We will see that it
provides us with a sensitive and non-trivial test of these models. In some cases, it
sheds new light on simple physical properties which were known before. In one case it
reveals an inconsistency in the literature and guides us towards its resolution.
Various special cases of Eqs.~(\ref{D4},\ref{D5}) have been discussed in the literature before \cite{3,4a}.

Throughout this paper we use units such that the physical fermion mass in the vacuum is $m=1$. 

{\em 1. DHN kink-antikink baryons}

In the massless GN model with discrete chiral symmetry $\psi \to \gamma_5 \psi$, [$m_0=0,\lambda=0$ in
Eq.~(\ref{D1})], Dashen, Hasslacher and Neveu (DHN) have constructed multi-fermion bound states
by inverse scattering theory \cite{5}. These states can be viewed as Hartree-Fock (HF) solutions belonging to the reflectionless
scalar potential 
\begin{equation}
S(x)=1+y(\tanh \xi_- - \tanh \xi_+)
\label{D6}
\end{equation}
with
\begin{equation}
\xi_{\pm}=yx \pm \frac{1}{2} {\rm artanh}\, y.
\label{D7}
\end{equation}
The Dirac spectrum with potential (\ref{D6}) is symmetric under $E\to -E$ (charge conjugation) and exhibits two
discrete states inside the mass gap at $E=\pm \sqrt{1-y^2}$, whereas the positive and negative energy continua
are depleted by one state each as compared to the vacuum. The negative energy continuum is fully occupied
($N$ fermions per level). Denoting the occupation numbers of the discrete states
by $n_{\pm}$, the self-consistency condition
\begin{equation}
\langle \bar{\psi}\psi \rangle = - \frac{1}{g^2} S(x)
\label{D8}
\end{equation}
relates the parameter $y$ to $n_{\pm}$ as follows,
\begin{equation}
y= \sin \theta, \quad \theta = \frac{\pi}{2N}(n_+-n_-+N).
\label{D9}
\end{equation}
The parameter $y$ controls the shape of $S(x)$ as well as the mass of the baryon, $M_B=2Ny/\pi$.
Let us now check Eqs.~(\ref{D4}) and (\ref{D5}).
Using the fermion single particle wave functions from Ref.~\cite{6} we
find that the expectation value of the current vanishes, $\langle j^1 \rangle=0$, as pointed out already
by Feinberg \cite{3}. This is consistent with Eq.~(\ref{D4}). 
Next we evaluate the fermion density (subtracting the density in the vacuum corresponding to $S(x)=1$),
\begin{equation}
\langle j^0  \rangle = \frac{N_f y}{4} \left( \frac{1}{\cosh^2 \xi_+}+\frac{1}{\cosh^2\xi_-}\right),
\label{D10}
\end{equation}
where $N_f$ is the fermion number 
\begin{equation}
N_f = \int {\rm d}x \langle j^0 \rangle =  n_++n_--N.
\label{D11}
\end{equation} 
The pseudoscalar condensate in the GN model is rarely looked at, since it enters neither the self-consistency condition
nor the calculation of the baryon mass.
If we evaluate it nevertheless in view of relation (\ref{D5}), we find
\begin{equation}
\langle \bar{\psi}{\rm i}\gamma_5 \psi \rangle = \frac{N_f y}{4} \left( \frac{1}{\cosh^2 \xi_+}-\frac{1}{\cosh^2\xi_-}\right).
\label{D12}
\end{equation}
The axial current equation (\ref{D5}) reduces to 
\begin{equation}
\partial_1 \langle j^0 \rangle = 2 S \langle \bar{\psi}{\rm i}\gamma_5 \psi \rangle
\label{D13}
\end{equation}
in the case at hand. (This relation was found earlier in Refs.~\cite{3,4a}.)
Inserting the bilinears from Eqs.~(\ref{D6},\ref{D10},\ref{D12}) into Eq.~(\ref{D13}) then yields the non-trivial identity
\begin{eqnarray}
& & y \partial_{\xi} \left(\frac{1}{\cosh^2 \xi_+} + \frac{1}{\cosh^2 \xi_-}\right)  = 
\label{D14} \\ 
& & 2 \left[1+y(\tanh \xi_--\tanh \xi_+)\right] \left(\frac{1}{\cosh^2 \xi_+} -\frac{1}{\cosh^2 \xi_-}\right)
\nonumber
\end{eqnarray}
which can indeed be verified for $\xi_{\pm}$ from Eq.~(\ref{D7}).

Knowing the pseudoscalar density is of interest for yet another reason. Any HF solution
of the GN model with $\langle \bar{\psi} {\rm i} \gamma_5 \psi \rangle =0$ is also a valid HF 
solution of the  
NJL$_2$ model. According to Eq.~(\ref{D12}), this is the case for all
DHN bound states with $N_f=0$,
i.e., ``baryonium" states. As can be seen from Eq.~(\ref{D11}) these states have
a total of $N$ fermions distributed over the two discrete levels (there are $N+1$ such states, including the vacuum)
and identically vanishing fermion density. 
In the massless NJL$_2$ model, the axial current is conserved and Eq.~(\ref{D5}) reduces to $\partial_1 \langle
j^0 \rangle=0$.
This is indeed satisfied by the GN baryonium states with vanishing $\langle
j^0 \rangle$. These particular solutions of the NJL$_2$ model have already been discussed by Shei \cite{7}. 

{\em 2. CCGZ kink baryons}

If $y \to 1$, the spatial extension of the DHN baryon becomes infinite and it decouples into two
separate entities. These can be regarded as another type of HF solution, the kink (or antikink)
with scalar potential 
\begin{equation}
S(x)= \pm \tanh x.
\label{D15}
\end{equation}
This topologically non-trivial object is known in the literature as Callan-Coleman-Gross-Zee (CCGZ) kink \cite{5}.
It has a single discrete state at zero energy which gets half of its strength from the positive and half from
the negative energy continua.
The fermion and pseudoscalar condensate are found to be ($\mp$ for kink and antikink),
\begin{eqnarray}
\langle j^0 \rangle &=&  \frac{N_f}{2 \cosh^2x},
\nonumber \\
\langle \bar{\psi} {\rm i}\gamma_5 \psi \rangle & = & \mp  \frac{N_f}{2 \cosh^2x},
\label{D16}
\end{eqnarray}
so that Eqs.~(\ref{D5},\ref{D13}) are again satisfied. The fermion number has the value 
\begin{equation}
N_f=n-\frac{N}{2}.
\label{D17}
\end{equation}
The kink is a celebrated example for fractional fermion number \cite{8}
(which we recover for odd $N$) and explains unusual spin-charge relations of solitons in polymers
like polyacetylene \cite{9}. 
Once again the baryonium state ($n=N/2,N_f=0$) has vanishing pseudoscalar condensate and fermion density,
solves the NJL$_2$ model and is consistent with axial current conservation. 

{\em 3. Shei baryons}

The massless NJL$_2$ model ($m_0=0,\lambda=1$) has continuous chiral symmetry $\psi\to \exp ({\rm i}\alpha \gamma_5) \psi$.
Here the mean field approach involves a scalar and a pseudoscalar potential with two self-consistency
conditions,
\begin{equation}
\langle \bar{\psi}\psi \rangle = - \frac{1}{g^2} S(x), \quad \langle \bar{\psi}{\rm i}\gamma_5 \psi \rangle =- \frac{1}{g^2} P(x).
\label{D18}
\end{equation}
Using the inverse scattering technique of DHN, Shei has identified massive baryons of this model \cite{7}. They
were later confirmed by Feinberg and Zee within a resolvent analysis of the Dirac operator \cite{10}. 
Since the HF formulation of these bound states is not yet available but will be needed below, we
have carried it out and included a brief summary in the appendix. In our notation, the self-consistent potentials of
the Shei baryons are 
\begin{eqnarray}
S(x)&=&1- f(x) \cos \theta , \quad P(x)\ =\ f(x) \sin \theta ,
\nonumber \\
f(x) & = & \frac{2\cos \theta}{1+\exp(2 \xi)},
\label{D19}
\end{eqnarray}
with $\xi=x \cos \theta$ and a parameter $\theta$ living in the interval [$ \pi/2,3\pi/2$].
There is a single discrete state at $E_0=-\sin \theta$ which crosses the full mass gap as $\theta$ is varied.
Filling all negative energy continuum states completely ($N$ fermions) and the discrete state partially ($n$ fermions),
selfconsistency demands
\begin{equation}
\theta= \left( \frac{3}{2}-\frac{n}{N}\right)\pi,
\label{D20}
\end{equation}
whereas the mass of the bound state can be shown to be
\begin{equation}
M_B= \frac{N}{\pi}\sin \frac{\pi n}{N}.
\label{D21}
\end{equation}
Now we compute the fermion density, using the single particle wave functions from the appendix. 
First of all we find that the positive energy continuum is depleted by $1-\nu$, the negative energy
continuum by $\nu$ in favor of the discrete state, with $\nu=n/N$. More specifically, the fermion
density from the negative energy continuum (after vacuum subtraction) is given by
\begin{equation}
\langle j^0 \rangle |_{\rm cont} =  \frac{n}{2} \frac{\cos \theta}{\cosh^2 \xi}  .
\label{D22}
\end{equation}
We recall that all continuum states are filled with $N$ fermions. The appearance of the factor $n$ is 
a combination of the filling factor $N$ and the depletion of the negative energy continuum states
by $n/N$.
The discrete state yields the density
\begin{equation}
\langle j^0 \rangle |_{\rm discr}  =  - \frac{n}{2} \frac{\cos \theta}{\cosh^2 \xi} , 
\label{D23}
\end{equation}
where now $n$ reflects directly the partial occupation of this state with $n$ fermions.
The total fermion density is the sum of both terms and vanishes identically, independently of $n$.
The induced fermion density and the valence fermion density cancel exactly, and $N_f=0$. 
Notice that the CCGZ kink with half filled discrete level is contained as a special case $\theta=\pi$
in the bound states found by Shei. As discussed above, it also has vanishing fermion number.

Total screening of fermion charge as a result of a cancellation between ``valence" and ``sea" fermions is
at first sight surprising. Going back to Eq.~(\ref{D5}), we recognize that the underlying reason is axial current conservation in the massless
NJL$_2$ model. It translates into the very restrictive condition $\partial_1 \langle j^0 \rangle=0$ for time
independent configurations.
Therefore, if the charge density vanishes at $|x| \to \infty$,
it has to be identically zero. Chiral symmetry forbids a local accumulation of fermions.
This effect is qualitatively reminiscent of the depletion of soliton charge by a finite chemical potential observed
in Refs.~\cite{11,12}, but it is more dramatic in the present case.

It may be worth pointing out that the Dirac-HF Hamiltonian with potentials (\ref{D19}),
after a global chiral rotation (see Eq.(\ref{A6}) from the appendix), coincides with the Hamiltonian
employed in the context of irrational
fractional fermion number \cite{13,14,15}. As shown in these works, the depletion of continuum states 
depends only on the asymptotics of the scalar potential and the value of the constant pseudoscalar
potential. In Ref.~\cite{15} dealing with relativistic field theory, the potentials were external fields without
any self-consistency condition
relating the occupation of the discrete level to the shape and size of the potential. In Ref.~\cite{14} on
diatomic polymers, self-consistency was only imposed on the scalar potential; the pseudoscalar one was a given
(charge conjugation symmetry violating) parameter. Hence in these works screening of charge
as discussed above was not an issue.

In the present case, unlike in the GN model, we are in conflict with the original assignment of baryon number by Shei. 
The baryon mass is symmetric around the midpoint $\theta=\pi$ of the interval in which $\theta$ lives, cf. Eq.~(\ref{A40})
of the appendix.
Shei interprets this symmetry as charge conjugation. Filling the valence level with 0 up to $N/2$ fermions is supposed
to describe baryons, from $N/2$ to $N$ antibaryons. We believe that the correct assignment is that 
all of the states found by Shei have $N_f=0$; they are no baryons at all, but rather baryonium states. 
A similar objection has been raised in the literature before by Farhi et al. \cite{16}
(see their appendix $C$), although no attempt was made to understand the vanishing baryon number. 
The same result can also be obtained from the resolvent analysis of the Dirac operator, see
the Appendix A of \cite{3} and Sect.~III of \cite{16a}.
Axial current conservation is a simple and strong argument in favor of these earlier works and the present study.
The lesson one learns here is that one cannot read off baryon number from the filling fraction of the 
valence level(s) in cases where the self-consistent potential induces fermion number.

{\em 4. Chiral spiral}

Another kind of baryon known in the NJL$_2$ model is the massless one due exclusively to induced fermion
number (``chiral spiral" with unit winding number \cite{17}). 
Here it is preferable to work in a finite box of length $L$. The relevant potentials 
\begin{equation}
S(x)=\cos \frac{2\pi Bx}{L}, \quad P(x)=\sin \frac{2\pi Bx}{L}
\label{D24}
\end{equation}
lie on the chiral circle $S^2+P^2=1$, i.e., the vacuum manifold
for spatially constant condensates. Since 
\begin{equation}
\gamma^0 S(x) +{\rm i}\gamma^1 P(x)= {\rm e}^{{\rm i}B\pi x \gamma_5/L} \gamma^0  {\rm e}^{-{\rm i}B\pi x\gamma_5/L},
\label{D25}
\end{equation}
one can map the HF problem with potentials (\ref{D24}) onto the vacuum problem with $S(x)=1,P(x)=0$ by a
local chiral rotation with linear $x$-dependence.
This induces spectral flow and increases the number of negative energy states by $B$, as can be shown either by
point splitting \cite{4a,17} or an appropriate cutoff method \cite{19}. 
The baryon density is constant (equal to $B/L$) so that the baryon is spread out over the whole space.
The baryon mass vanishes in the limit $L\to \infty$. These massless baryons \ -- also known from the 't Hooft model 
\cite{20,21} --\  fit nicely into the present discussion,
illustrating the fact that continuous chiral symmetry enforces spatially constant fermion density.
So far, only integer baryon number was considered ($B=1$ corresponds to $N$ extra fermions, so that $B=N_f/N$).
Presumably one could also allow for values $N_f<N$ by letting the potential make only a fraction of a whole turn. 
This possibility was not envisaged so far, since it is most likely not relevant for dense matter and the phase diagram
of the model.

In summary, it seems that the chiral spiral type baryons are the only multifermion bound states of the
massless NJL$_2$ model which carry fermion number. They are completely delocalized as a result of axial
current conservation.
\newpage
{\em 5. Phase diagram of massless NJL$_2$ model}

In Ref.~\cite{2} the phase diagram of the massless NJL$_2$ model at finite temperature $T$ and chemical potential $\mu$
was derived. The HF solution again exhibits a potential of helical shape (chiral spiral), with radius depending only
on temperature and pitch depending only on chemical potential. In the present context, the most
important finding is the fact that the fermion density is $x$-independent everywhere in the ($T,\mu$) plane. 
The breakdown of translational invariance as a result of the Peierls instability (gap formation
at the Fermi surface) only shows up in the (periodic) scalar and pseudoscalar condensates. These findings are
consistent with Eq.~(\ref{D5}) which predicts $\partial_1 \langle j^0 \rangle=0$ for the grand canonical 
expectation value of the density. The reason behind the surprisingly simple phase diagram (as compared, 
e.g., to the GN model with discrete chiral symmetry \cite{4}) then becomes clearer.

{\em 6. Bare fermion mass corrections}

If we switch on the bare mass $m_0$ in the GN model, Eq.~(\ref{D13}) remains valid with the scalar potential
now given by $S=m_0-g^2 \langle \bar{\psi}\psi \rangle$. The discussion of the kink-antikink baryons can be repeated
literally, the only change being the relation between $y$ and the occupation numbers $n_{\pm}$. 
The topological kink does not exist anymore because the degeneracy between the two vacua in the massless GN
model is lifted by the bare mass term.

Turning on $m_0$ in the NJL$_2$ model is more interesting. Axial current conservation (\ref{D5}) gets replaced by the PCAC
relation 
\begin{equation}
\partial_1 \langle j^0 \rangle = 2 m_0 \langle \bar{\psi}{\rm i}\gamma_5 \psi\rangle = - \frac{2N\gamma}{\pi} P,
\label{D26}
\end{equation}
where we have used $\langle \bar{\psi}{\rm i}\gamma_5 \psi \rangle =-P/g^2$ and expressed the ratio of the two bare
parameters $m_0$ and $g^2$ by the physical parameter
\begin{equation}
\gamma = \frac{\pi m_0}{Ng^2}.
\label{D27}
\end{equation}
It is related to the ``pion" mass via \cite{22}
\begin{equation}
\gamma= \frac{1}{\sqrt{\eta-1}} \arctan \frac{1}{\sqrt{\eta-1}}, \quad \eta=\frac{4}{m_{\pi}^2}.
\label{D28}
\end{equation}
In Ref.~\cite{23} the baryon in the massive NJL$_2$ model has been determined approximately
by means of the derivative expansion, leading to a chiral perturbation series in powers of $m_{\pi}^2$.
To lowest order, the fermion density and the pseudoscalar condensate are given by
\begin{equation}
\langle j^0 \rangle  = \frac{N}{\pi} \partial_x \chi, \quad P=-\sin 2 \chi
\label{D29}
\end{equation}
where $2\chi=4 \arctan {\rm e}^{m_{\pi}x}$ is the sine-Gordon kink.
Inserting Eqs.~(\ref{D29}) into Eq.~(\ref{D26}) and using the leading order approximation $\gamma \approx m_{\pi}^2/4$
to Eq.~(\ref{D28}), we recover indeed the static sine-Gordon equation for the chiral phase ($\xi=m_{\pi}x$),
\begin{equation}
\partial_\xi^2 (2 \chi) = \sin (2\chi).
\label{D30}
\end{equation}
It was already obtained in Ref.~\cite{4a} and systematically improved by means of the derivative expansion in Ref.~\cite{23}.
Turning now to the higher order corrections, we integrate Eq.~(\ref{D26}) to obtain the fermion density,
\begin{equation}
\langle j^0(x) \rangle  = -\frac{2N\gamma}{\pi} \int_{-\infty}^x {\rm d}x' P(x') .
\label{D31}
\end{equation} 
Another integration over $x$ yields the fermion number $N_f=N$ and the ``sum rule"
\begin{equation}
1 = \frac{2\gamma}{\pi} \int_{-\infty}^{\infty}{\rm d}x  x P(x)
\label{D32}
\end{equation} 
where we have integrated by parts and used $P(-x)=-P(x)$. This relation provides us with a new way of testing the results
of the derivative expansion. Inserting $P(x)$ from Ref.~\cite{23} and $\gamma$ from Eq.~(\ref{D28}),
we find indeed analytically that the right hand side of Eq.~(\ref{D31}) is 1+O($m_{\pi}^8$), consistent with the order
of the calculation. We have also checked that the higher order results of Urlichs \cite{24}
reduce the error to O($m_{\pi}^{14}$), an excellent test of a lengthy and complicated analytical calculation. 
Relation (\ref{D26}) should also hold at finite temperature and chemical potential
and may be of some help in determining the (yet largely unknown) phase diagram of the massive NJL$_2$ model \cite{25}.

Let us summarize our results. The most important equation of this paper is Eq.~(\ref{D5}). It relates
the spatial derivative of the fermion density to the bare mass and scalar/pseudoscalar condensates
in static configurations of GN models. As the derivation in a few lines shows, it is almost a triviality.
Nevertheless it has many implications some of which have been exploited before \cite{4a}.
The most striking predictions arise in the case where axial current
is conserved, i.e., the massless NJL$_2$ model. Here the combination of axial current conservation
and the close relationship between vector and axial currents in 1+1 dimensions constrains the theory most severely.
As a result the fermion density is spatially constant in any static configuration of the massless NJL$_2$ model,
in agreement with previous calculations of massless baryons and the phase diagram based on the chiral
spiral. Our result is in conflict with the Shei baryon, but this can be resolved
by realizing that fermion density actually vanishes as a result of a cancellation between valence and sea
fermions. In those cases where the right hand side of Eq.~(\ref{D5}) does not vanish, the identity
has proven quite powerful for testing known results. We applied it successfully to the kink and kink-antikink baryons
in the GN model, as well as to finite bare mass corrections to baryons in the massive NJL$_2$ model.
In all of these cases we confirmed the published results. Perhaps most noteworthy is the ``sum rule" (\ref{D32})
relating the first moment of the pseudoscalar density to baryon number. It enabled us to test
rather involved analytic calculations based on the derivative expansion to high order in a non-trivial way.
\newpage
{\bf Acknowledgement:}

We should like to thank Gerald Dunne, Joshua Feinberg and Herbert Weigel for useful correspondence concerning
Refs.~\cite{7,10,16}.
\vskip 0.5cm
%################################################################################################
%                                                                                           APPENDIX
%################################################################################################
{\bf Appendix: Hartree-Fock treatment of Shei bound state}
\vskip 0.5cm
Starting point is the Dirac-HF equation with the potential from Refs.~\cite{7,10},
\begin{equation}
\left( \gamma_5 \frac{1}{\rm i}\partial_x + \gamma^0 (1-f(x)\cos \theta) + {\rm i}\gamma^1
f(x)\sin \theta \right) \psi = E \psi
\label{A1}
\end{equation}
where
\begin{equation}
f(x)=\frac{2 k_0}{1+\exp(2\xi)}, \quad \xi=k_0 x, \quad k_0=\cos \theta.
\label{A2}
\end{equation}
The global chiral transformation
\begin{equation}
\psi \to {\rm e}^{{\rm i} \theta \gamma_5/2} \psi
\label{A3}
\end{equation}
maps the pseudoscalar potential onto a constant,
\begin{equation}
\left( \gamma_5 \frac{1}{\rm i}\partial_x + \gamma^0 (\cos \theta-f(x)) + {\rm i}\gamma^1
\sin \theta \right) \psi = E \psi.
\label{A4}
\end{equation}
Using the representation
\begin{equation}
\gamma^0 = - \sigma_1,\quad \gamma^1 = {\rm i} \sigma_3, \quad \gamma_5 = - \sigma_2,
\label{A5}
\end{equation}
the Hamiltonian now reads
\begin{equation}
H = \left( \begin{array}{cc} -\sin \theta & \partial_x - \cos \theta + f(x) \\ - \partial_x - \cos \theta + f(x) &
\sin \theta \end{array} \right)
\label{A6}
\end{equation}
and becomes diagonal in the Dirac indices upon squaring,
\begin{equation}
H^2 = \left( \begin{array}{cc} - \partial_x^2  + U_+  & 0 \\
0 & - \partial_x^2 + U_- \end{array} \right)
\label{A7}
\end{equation}
with 
\begin{equation}
U_{\pm} = f^2 \pm f' +1-2f\cos \theta.
\label{A7a}
\end{equation}
The resulting Schr\"odinger-type equations for the upper and lower components $\psi_{\pm}$ of $\psi$ are identical to those of the kink
in the GN model,
\begin{eqnarray}
\left(  \partial_{\xi}^2 + \frac{2}{\cosh^2\xi} \right) \psi_+ &=& - \kappa^2 \psi_+ 
\nonumber \\
 \partial_{\xi}^2 \psi_- & = &  - \kappa^2  \psi_- 
\label{A8}
\end{eqnarray}
with 
\begin{equation}
\kappa^2 = \frac{E^2-1}{k_0^2}.
\label{A9}
\end{equation}
There is one discrete state,
\begin{equation}
\psi_0 = \sqrt{\frac{|k_0|}{2}} \left( \begin{array}{c} \frac{1}{\cosh{\xi}} \\ 0 \end{array} \right), \quad E_0=- \sin \theta,
\label{A10}
\end{equation}
whereas normalized positive and negative energy continuum states are given by ($E=\pm \sqrt{k^2+1}$)
\begin{equation}
\psi_E(x) = \frac{1}{\sqrt{2E(E-E_0)}} \left( \begin{array}{c} {\rm i}k-k_0 \tanh \xi \\ E - E_0 \end{array} \right) {\rm e}^{{\rm i}kx}.
\label{A11}
\end{equation}
We note the following 
contributions of single particle states to the condensates,
\begin{eqnarray}
\bar{\psi}_0 \psi_0 & = & 0 
\nonumber \\
\bar{\psi}_0 {\rm i} \gamma_5 \psi_0 & =&  - \frac{|k_0|}{2} \frac{1}{\cosh^2 \xi}
\nonumber \\
\bar{\psi}_E \psi_E & = & \frac{k_0 \tanh \xi}{E}
\label{A12}   \\
\bar{\psi}_E {\rm i} \gamma_5 \psi_E & = & \frac{k_0^2}{2E_0 \cosh^2\xi} \left(\frac{1}{E-E_0}- \frac{1}{E}\right) - \frac{E_0}{E}
\nonumber
\end{eqnarray}
Self-consistency of the scalar potential
\begin{equation}
\cos \theta - f(x) = - N g^2 \int_{- \Lambda/2}^{\Lambda/2} \frac{{\rm d}k}{2\pi} \bar{\psi}_k^{(-)}\psi_k^{(-)} 
\label{A13}
\end{equation}
can be verified with the help of the vacuum gap equation $(Ng^2/\pi) \ln \Lambda=1$. The self-consistency 
condition for the pseudoscalar potential, assuming $n$ fermions in the discrete state, reads
\begin{equation}
\sin \theta  = -ng^2  \bar{\psi}_0 {\rm i} \gamma_5 \psi_0 -Ng^2 \int_{- \Lambda/2}^{\Lambda/2}
\frac{{\rm d}k}{2\pi} \bar{\psi}_k^{(-)}{\rm i}\gamma_5 \psi_k^{(-)}.
\label{A14}
\end{equation} 
It holds for the following relation between $\theta$
and $\nu=n/N$,
\begin{equation}
\nu = - \frac{2}{\pi} \arctan \left(\frac{1+\sin \theta}{\cos \theta}\right), \
\theta= \left( \frac{3}{2} - \nu \right) \pi .
\label{A15}
\end{equation}  
If $\nu$ runs from 0 to 1, $\theta$ runs from $3\pi/2$ to $\pi/2$ and $E_0$ from 1 to -1, thus the discrete
level crosses the full mass gap from top to bottom.
The fermion density for discrete and continuum states is 
\begin{eqnarray}
\psi_0^{\dagger}\psi_0 &=& \frac{|k_0|}{2 \cosh^2 \xi},
\nonumber \\
\psi_E^{\dagger}\psi_E & = & 1- \frac{k_0^2}{2E(E-E_0)\cosh^2 \xi}.
\label{A16}
\end{eqnarray}
After subtracting the 1 in $\psi_E^{\dagger} \psi_E$,
this was used to evaluate the fermion densities and continuum depletion factors in the main text.

The solution discussed so far is related to the kink of the GN model. We now briefly sketch the
construction of the antikink needed for evaluating the mass of the bound state below.
Here, we start from the ansatz
\begin{equation}
S(x)=1- \tilde{f}(x) \cos \theta, \quad P(x)= \tilde{f}(x) \sin \theta 
\label{A17}
\end{equation}
where now
\begin{equation}
\tilde{f}(x) = 2 \cos \theta - f(x).
\label{A18}
\end{equation}
The relationship between $f$ and $\tilde{f}$ becomes more transparent upon writing
\begin{eqnarray}
f(x)&=& \cos \theta (1-\tanh \xi),
\nonumber \\
\tilde{f}(x) & = & \cos \theta (1+\tanh \xi).
\label{A19}
\end{eqnarray}
Everything goes through as before except that $f$ has to be replaced by $\tilde{f}$.
In Eqs.~(\ref{A7},\ref{A8}), this leads to an interchange of upper and lower components.
We now find the discrete state,
\begin{equation}
\psi_0 = \sqrt{\frac{|k_0|}{2}} \left( \begin{array}{c} 0 \\ \frac{1}{\cosh{\xi}} \end{array} \right), \quad \tilde{E}_0= \sin \theta,
\label{A20}
\end{equation}
and the continuum states ($E=\pm \sqrt{k^2+1}$), 
\begin{equation}
\psi_E(x) = \frac{1}{\sqrt{2E(E-\tilde{E}_0)}} \left( \begin{array}{c} -(E-\tilde{E}_0) \\ {\rm i}k-k_0 \tanh \xi \end{array} \right) {\rm e}^{{\rm i}kx}.
\label{A21}
\end{equation}
The relation between the kink and antikink fermion wave functions is a discrete $\gamma_5$-transformation 
combined with the reflection $\theta \to 2\pi-\theta$.
Assuming $\tilde{n}$ fermions in the discrete state (occupation fraction
$\tilde{\nu}=\tilde{n}/N$), condition (\ref{A15}) gets replaced by  
\begin{equation}
\tilde{\nu} = - \frac{2}{\pi} \arctan \left(\frac{1-\sin \theta}{\cos \theta}\right), \ \theta= \left( \frac{1}{2}+ \tilde{\nu} \right) \pi .
\label{A22}
\end{equation}
Notice that $\nu+\tilde{\nu} = 1$ for a common chiral angle $\theta$ for kink and antikink.

We are now in a position to evaluate the mass of the bound state.
Since the kink and the vacuum cannot be described with the same boundary conditions
(there is a chiral twist in the kink), it is not possible to apply directly the method from the GN model \cite{5,6}.
Therefore we simply glue a kink and an antikink together, compute the mass of this whole
object with periodic boundary conditions (the antikink un-does the chiral twist) and divide by two.

For this purpose we need the sum over single particle energies from the discrete states, the HF double counting correction
and the integral over (negative energy) continuum single particle energies. The discrete states give
\begin{equation}
\Delta M_B|_{\rm discr} = n E_0 + \tilde{n} \tilde{E}_0 = N(1-2\nu)\sin \theta.
\label{A23}
\end{equation}
The double counting correction is (using the gap equation)
\begin{eqnarray}
\Delta M_B|_{\rm dc} & = & \frac{1}{g^2} \int {\rm d}x (S^2+P^2-1)
\nonumber \\
& = & - \frac{k_0^2}{g^2} \int {\rm d}x \frac{1}{\cosh^2 \xi}
\nonumber \\
& = & \frac{2N}{\pi} \cos \theta \ln \Lambda.
\label{A24}
\end{eqnarray}
For the continuum states we proceed as follows:
Negative energy continuum spinors for the kink evolve as 
follows from large negative to large positive $x$ ($E_k=\sqrt{k^2+1}$),
\begin{eqnarray}
& & \frac{1}{\sqrt{2E_k(E_k-\sin \theta)}}\left( \begin{array}{c} {\rm i}k- \cos \theta \\
-E_k+ \sin \theta \end{array} \right) \to 
\nonumber \\
& &
\frac{1}{\sqrt{2E_k(E_k-\sin \theta)}}\left( \begin{array}{c} {\rm i}k+ \cos \theta \\
-E_k+ \sin \theta \end{array} \right).
\label{A25}
\end{eqnarray}
For the antikink, the corresponding relation is 
\begin{eqnarray}
& & \frac{1}{\sqrt{2E_k(E_k+\sin \theta)}}\left( \begin{array}{c} E_k + \sin \theta \\ {\rm i}k- \cos \theta 
\end{array} \right) \to
\nonumber \\
& &  \frac{1}{\sqrt{2E_k(E_k+\sin \theta)}}\left( \begin{array}{c} E_k+\sin \theta \\ {\rm i}k+ \cos \theta 
\end{array} \right).
\label{A26}
\end{eqnarray}
When glueing the two objects together, we therefore have to multiply the antikink spinor by the phase factor
\begin{equation}
C = \frac{{\rm i}k + \cos \theta}{\sqrt{k^2+\cos^2\theta}}.
\label{A27}
\end{equation}
This maps the antikink spinor at large negative $x$ onto the kink spinor at large positive $x$. We then
find that the kink-antikink system produces the total phase shift
\begin{equation}
\delta(k) = - \arctan \frac{2 k \cos \theta}{k^2-\cos^2 \theta}.
\label{A28}
\end{equation}
This is common to upper and lower components, i.e., there is no chiral twist anymore.
As a consequence the evaluation of $\Delta M_B|_{\rm cont}$ is now straightforward \cite{5,6},
\begin{eqnarray}
\Delta M_B|_{\rm cont} & = & N \int \frac{{\rm d}k}{2\pi} \delta(k) \frac{{\rm d}E_k}{{\rm d}k}
\nonumber \\
& = & \frac{N}{2\pi} \delta(k)E_k|_{-\infty}^{\infty} - N \int_{-\Lambda/2}^{\Lambda/2} \frac{{\rm d}k}{2\pi} \frac{{\rm d}\delta}
{{\rm d}k} E_k
\nonumber \\
& = & \Delta M_B|_{\rm surf} + \Delta M_B|_{\rm bulk}.
\label{A29}
\end{eqnarray}
Using the asymptotics
\begin{equation}
\delta \approx - \frac{2\cos \theta}{k} \quad (k \to \pm \infty)
\label{A30}
\end{equation}
and the derivative
\begin{equation}
\frac{{\rm d}\delta}{{\rm d}k} = \frac{2 \cos \theta}{k^2+\cos^2 \theta},
\label{A31}
\end{equation}
we find
\begin{eqnarray}
\Delta M_B|_{\rm surf} &=&  - \frac{2N}{\pi} \cos \theta ,
\label{A32} \\
\Delta M_B|_{\rm bulk} &=& - \frac{2N}{\pi} \cos \theta \ln \Lambda  -N(1-2\nu) \sin \theta.
\nonumber
\end{eqnarray}
The first part of $\Delta M_B|_{\rm bulk}$ cancels the double counting correction (\ref{A24}), the second part cancels the discrete state
contribution (\ref{A23}) so that only $\Delta M_B|_{\rm surf}$ survives. Dividing by 2 we finally get
\begin{equation}
M_B=- \frac{N}{\pi}\cos \theta = \frac{N}{\pi} \sin \pi \nu = \frac{N}{\pi} \sin \pi \tilde{\nu}
\label{A40}
\end{equation}
in agreement with Refs.~\cite{7,10}.

%##################################################################

\end{document}